# Evaluating Quality of Chatbots
# and Intelligent Conversational Agents

Nicole Radziwill and Morgan Benton

**Abstract**: Chatbots are one class of intelligent, conversational software agents activated by natural language input (which can be in the form of text, voice, or both). They provide conversational output in response, and if commanded, can sometimes also execute tasks. Although chatbot technologies have existed since the 1960's and have influenced user interface development in games since the early 1980's, chatbots are now easier to train and implement. This is due to plentiful open source code, widely available development platforms, and implementation options via Software as a Service (SaaS). In addition to enhancing customer experiences and supporting learning, chatbots can also be used to engineer social harm - that is, to spread rumors and misinformation, or attack people for posting their thoughts and opinions online. This paper presents a literature review of quality issues and attributes as they relate to the contemporary issue of chatbot development and implementation. Finally, quality assessment approaches are reviewed, and a quality assessment method based on these attributes and the Analytic Hierarchy Process (AHP) is proposed and examined.

**Keywords**: chatbot, artificial intelligence, intelligent agents, usability, quality attributes, Analytic Hierarchy Process (AHP)

## Introduction

> *"The most profound technologies are those that disappear. They weave themselves into the fabric of everyday life until they are indistinguishable from it." - Weiser (1991)*

Whether you realize it or not, the "people" you interact with online are not all people. Customer service chat and commercial social media interactions are increasingly managed by intelligent agents, many of which have been developed with human identities and even personalities. (Simonite, 2017) Even though the technology itself is not new, reliable linguistic functionality, availability through Software as a Service (SaaS), and the addition of intelligence through machine learning has increased its popularity. Between 2007 and 2015, chatbots were participating in a third to a half of all online interactions (Tsvetkova et al., 2016) and the rate at which new chatbots are being deployed has increased since then.

Social, conversational bots can be used to provide benefits to companies, who use them to reduce time-to-response, provide enhanced customer service, increase satisfaction, and increase engagement. Unfortunately, some chatbots are specifically designed to be harmful. For example, networks of fake users (called "sybils" on Twitter) have been implemented to artificially inflate "follower" counts to increase social status for users who purchase them, to spread fake news or rumors, and even to intimidate users who express certain political beliefs. (Ferrara et al., 2016)

In the 2016 US Presidential election, up to a fifth of the comments and responses on Twitter were driven by fully or partially autonomous Twitter accounts. (Albright, 2017)

Due to their flexibility and ease of use (compared to web-based or mobile applications), some have speculated that conversational agents may be a universal user interface and may replace "apps" - a democratization of the versatile command line. (Solomon, 2017) In addition, chatbots and conversational agents are anticipated to be important interfaces in Virtual Reality (VR) environments. (Beilby & Zakos, 2014) As a result, it is important to understand the issues and quality attributes associated with developing and implementing high-quality chatbots and conversational agents, and identify a mechanism for quality assurance across these factors.

The rest of this paper is organized as follows. First, we present the historical context of chatbots and the background for chatbots and conversational agents, which are two types of dialog systems. Next, we outline the methodology for our literature review along two topics: 1) quality attributes for chatbots and conversational agents, and 2) quality assessment approaches. Finally, we synthesize these results and present an approach for evaluating the quality of these technologies in terms of key quality attributes.

**Background**

Chatbot systems originated with programs like ELIZA (Weizenbaum, 1966) which were intended to demonstrate natural language conversation with a computer. An early stated goal of such systems was to pass the Turing Test (Turing, 1950), in which a human interrogator deems a computer sufficiently "intelligent" to pass as a human. However, primitive systems like ELIZA used keyword matching and minimal context identification, and lacked the ability to keep a conversation going. As shown in Figure 1, through interactions with the program, it was easy to guess that ELIZA was a computer.

Researchers continued to develop demonstration systems with natural language capabilities, but none were capable of passing the Turing Test. In the early 1980's, ALICE (Wallace, 2003) was created, becoming significant not for its conversational capabilities but because it led to the the development of Artificial Intelligence Markup Language (AIML). AIML is used to declare pattern-matching rules that links user-submitted words and phrases with topic categories. It is eXtensible Markup Language (XML) based, and supports most chatbot platforms and services in use today.

```
                EEEEEE  LL      IIII  ZZZZZZZ  AAAAA
                EE      LL       II        ZZ  AA   AA
                EEEEE   LL       II       ZZZ  AAAAAAA
                EE      LL       II       ZZ   AA   AA
                EEEEEE  LLLLLL  IIII  ZZZZZZZ  AA   AA

  Eliza is a mock Rogerian psychotherapist.
  The original program was described by Joseph Weizenbaum in 1966.
  This implementation by Norbert Landsteiner 2005.

ELIZA: Is something troubling you ?
YOU:   Yes. I'm trying to describe you for a research paper I'm writing. What sh
ould I say?
ELIZA: You are sure.
YOU:   Yes, I'm sure.
ELIZA: I see.
YOU:   What can I tell them about you?
ELIZA: You seem to need to tell them about me.
YOU:   I absolutely need to. Please help.
ELIZA: That's quite interesting.
YOU:   I agree.
ELIZA: You say you agree ?
YOU:
```

**Figure 1.** A sample dialog with ELIZA (from http://www.masswerk.at/elizabot/eliza.html).

Chatbots receive natural language input, sometimes interpreted through speech recognition software, and execute one or more related commands to engage in goal-directed behavior (often on behalf of a human user). As intelligent agents, they are usually autonomous, reactive, proactive, and social. The most advanced systems employ machine learning (often Markov chains or deep neural networks) so that they may also adapt to new information or new requests.

Chatbots are one category of conversational agents, which are software systems that mimic interactions with real people. They are typically not embodied in the forms of animals, avatars, humans, or humanoid robots (those programs are considered to be "embodied conversational agents"). Conversational agents, a class of dialog systems, have been a subject of research in communications for decades. Interactive Voice Response (IVR) systems (e.g. "Press or Say 1 for English") are also dialog systems, but are not usually considered conversational agents since they implement decision trees. (McTear et al., 2016) These terms are related to each other in Figure 2.

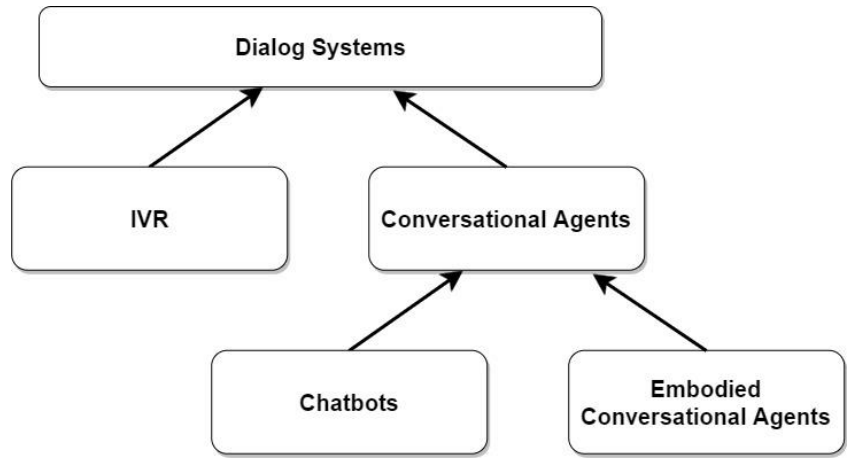

**Figure 2.** Relationships between classes of software-based dialog systems.

The emphasis in this paper is on the text-based conversational agents found online and in some Internet of Things (IoT) devices, which are sometimes (but not often) embodied. This contrasts with voice-activated conversational agents like Siri, Google Now, Cortana, Samsung S Voice, which are not considered chatbots. Although the earliest software in this genre appeared in Internet Relay Chat (IRC) environments in the mid-1990's, bots have evolved to serve multiple purposes, such as content editing and brokering complex transactions. (Tsvetkova et al., 2016) These programs serve a range of roles, from personal assistant, to intelligent virtual agent, to companion. There are agents designed to serve as personal university advisors (Ghose & Barua, 2013), educational agents developed to help improve learning outcomes (Kerly et al., 2007), and "Art-Bots" (Vassos et al., 2016) to engage museum visitors in participatory installations.

Chatbots are distinct from *bots*, compromised computers that often run malicious software and can be linked together as *botnets* to coordinate large-scale denial of service attacks (Thing et al., 2007). However, chatbots can be launched from botnets to shape social perceptions. No one who spends time online is immune from the potential harm of chatbots. Even the director of the annual Loebner Prize Competition in Artificial Intelligence, an event that pits the most sophisticated chatbots against one another, was fooled into thinking a chatbot on a dating service was interested in him romantically. (Epstein, 2007)

Researchers are actively investigating ways to mitigate the harm from chatbots that engage in social engineering. Alarifi et al. (2016), for example, built a corpus of these "sybils" and a browser plug-in to help human users better distinguish between humans and machines. The harm derives from the ease with which these chatbots, especially on social networks like Twitter, pass the Turing Test and convince human participants that their criticisms, abuse, and even rape and death threats are real and originate with other humans. (McElrath, 2017) The potential for social engineering emphasizes the need to critically examine quality attributes for chatbots, in part to protect the well-being of individuals and societies.

Development and implementation of chatbots today is easier, and chatbots themselves are more powerful. Development platforms, some of which implement Software as a Service (SaaS) approaches (e.g. Pandorabots, Chatfuel, Botsify) serve to split the testing responsibility between service provider (who is responsible for testing inputs, execution of actions, and realistic outputs) and client (who evaluates ease of use and effectiveness of task accomplishment). Because of easy integration with social media and developer productivity tools (e.g. Slack, GitHub) chatbots may even be instrumental in improving the work processes in traditional and agile development teams. (Storey & Zagalsky, 2016)

Machine learning approaches are also increasingly integrated to make these agents more adaptive to different input styles and new tasks. Systems are no longer dependent on deterministic responses from rules-based pattern matching, like ELIZA and ALICE. More commonly, systems leverage supervised learning (which requires large training sets), unsupervised learning (like Markov-chain based models), and hybrid intelligence (where humans participate in the training process over time). Supervised learning and hybrid intelligence approaches are more extensive and costly, but can result in systems that are better at on-the-fly problem solving and take less time to achieve goals. (Wilson et al., 2017)

**Methodology**

The purpose of this research was to examine the academic and industry literature to 1) provide a comprehensive review of quality attributes for chatbots and conversational agents, and 2) identify appropriate quality assurance approaches. To do this, publications on quality in chatbots and conversational agents were identified through systematic searches of Google Scholar, JSTOR, and EBSCO Host from 1990 to 2017. The search terms that were used were 'chatbots', 'conversational agents', 'embodied conversational agents', 'quality', 'quality attributes', 'quality of chatbots', 'quality of conversational agents', and 'quality assurance' in various combinations. Papers were drawn from the domains of engineering, technology, psychology, communications, and anthropology.

The selection of articles was based on three criteria. First, scholarly references were emphasized, supplemented only by industry publications from 2016 and 2017. Publications were selected if they contained at least one of the search terms in the title and/or abstract to ensure the relevance of data collection. Only publications were selected that had quality as a major or influential aspect of the research. Finally, highly technical articles focused on programming and engineering aspects of chatbots, including improving the quality of speech recognition, were excluded.

The initial search generated a sampling frame of 7,340 articles, which was further refined by adding search terms for 'evaluation', 'assessment', 'quality metrics' and 'metrics'. Most recent articles (from 2016 and 2017) were inspected next, followed by articles between 2013 and 2015, and then from 2007 to 2012. The selection of time frames was made to limit the number of articles for review to less than 300, and ultimately, 36 scholarly articles and conference papers were determined to be relevant to the objectives of this paper. They were supplemented with 10 articles from industry or trade magazines. Beyond this, only seven articles were identified for the second portion of the study focusing on quality assurance, and all were used in the analysis.

**Quality Attributes**

We extracted quality attributes from each of the 32 papers and 10 articles, and grouped them based on similarity. After two or three iterations, we noticed that in general, they were aligned with the ISO 9241 concept of usability: "The **effectiveness**, **efficiency** and **satisfaction** with which specified users achieve specified goals in particular environments." (Abran et al., 2003) In particular, effectiveness refers to the accuracy and completeness with which specified users achieve their goals, and efficiency refers to how well resources are applied to achieve those goals. Not surprisingly, particularly in a SaaS environment, the burden of demonstrating efficiency and effectiveness falls more on the service provider (since this will rely on common functionality), while the need to ensure that customers are satisfied will remain with the implementer. Table 1 outlines the quality attributes organized in terms of ISO 9241.

| EFFICIENCY | | |
|---|---|---|
| **Category** | **Quality Attribute** | **Reference** |
| Performance | <ul><li>Graceful degradation</li><li>Robustness to manipulation</li><li>Robustness to unexpected input</li><li>Avoid inappropriate utterances and be able to perform damage control</li><li>Effective function allocation, provides appropriate escalation channels to humans</li></ul> | <ul><li>Cohen & Lane (2016)</li><li>Thieltges (2016)</li><li>Kluwer (2011)</li><li>Morrissey and Kirakowski (2013)</li><li>Staven (2017)</li></ul> |
| **EFFECTIVENESS** | | |
| **Category** | **Quality Attribute** | **Reference** |
| Functionality | <ul><li>Accurate speech synthesis</li><li>Interprets commands accurately</li><li>Use appropriate degrees of formality, linguistic register</li><li>Linguistic accuracy of outputs</li><li>Execute requested tasks</li><li>Facilitate transactions and follows up with status reports</li><li>General ease of use</li><li>Engage in on-the-fly problem solving</li><li>Contains breadth of knowledge, is flexible in interpreting it</li></ul> | <ul><li>Kuligowska (2015)</li><li>Eeuwen (2017)</li><li>Morrissey & Kirakowski (2013)</li><li>Wallace (2003)</li><li>Ramos (2017)</li><li>Eeuwen (2017)</li><li>Solomon (2017)</li><li>Cohen & Lane (2016)</li></ul> |
| Humanity | <ul><li>Passes the Turing test</li></ul> | <ul><li>Weizenbaum (1966); Wallace (2003)</li></ul> |
| | ● Does not have to pass the Turing Test<br>● Transparent to inspection, discloses its chatbot identity<br>● Include errors to increase realism<br>● Convincing, satisfying, & natural interaction<br>● Able to respond to specific questions<br>● Able to maintain themed discussion | ● Ramos (2017)<br>● Bostrom & Yudkowski (2014)<br>● Coniam (2014)<br>● Morrissey & Kirakowski (2013) |
| **SATISFACTION** | | |
| **Category** | **Quality Attribute** | **Reference** |
| Affect | ● Provide greetings, convey personality<br>● Give conversational cues<br>● Provide emotional information through tone, inflection, and expressivity<br>● Exude warmth and authenticity<br>● Make tasks more fun and interesting<br>● Entertain and/or enable participant to enjoy the interaction<br>● Read and respond to moods of human participant | ● Morrissey & Kirakowski (2013)<br>● Pauletto et al. (2013)<br>● Solomon (2017)<br>● Eeuwen (2017)<br>● Ramos (2017)<br>● Meira & Canuto (2015) |
| Ethics & Behavior | ● Respect, inclusion, and preservation of dignity (linked to choice of training set)<br>● Ethics and cultural knowledge of users<br>● Protect and respect privacy<br>● Nondeception<br>● Sensitivity to safety and social concerns<br>● Trustworthiness (linked to perceived quality)<br>● Awareness of trends and social context | ● Neff & Nagy (2016)<br>● Applin & Fischer (2015)<br>● Eeuwen (2017)<br>● Isaac & Bridewell (2014)<br>● Miner et al. (2016)<br>● Herzum et al. (2002)<br>● Vetter (2002) |
| Accessibility | ● Responds to social cues or lack thereof<br>● Can detect meaning or intent<br>● Meets neurodiverse needs such as extra response time and text interface | ● Morrissey and Kirakowski (2013)<br>● Wilson et al. (2017)<br>● Radziwill & Benton (2017) |

**Table 1.** Quality attributes of chatbots and conversational agents.

There was remarkable consistency across the source material regarding quality attributes; although several of them were repeated across sources, only the primary source (or the source where that attribute was emphasized) appear in Table 1. However, one attribute was in question: whether or not a chatbot or conversational agent should pass the Turing Test. According to most researchers, and following the earliest conversational interfaces, responding and interacting like a human should be the top priority. In fact, this principle guided development for nearly four decades since ELIZA came online.

According to Ramos (2017), who references the Facebook chatbot (named Joy) in addition to a character from a 2014 Disney movie, argues that giving the impression of being human is not a valid quality attribute. "When interacting with a fictional character," he claims, "people are willing to have a suspension of disbelief and enjoy the interaction. In our user testing, I've seen people interact with Joy. When they get a response with an emoji, I've seen the smiles. People know it's not real and don't care that it wouldn't pass a Turing test — especially when Joy saves them money. Just think of the character Baymax from the movie Big Hero Six: He doesn't sound or look human, but that doesn't prevent you from empathizing with him. Would Baymax pass a Turing test? No, but that's not what matters."

The position that a chatbot or conversational agent should *not* aim to act human is shared by some analysts like Wilson et al. (2017) who believe that character development and character management is one of the new jobs that will be enabled by the spread of artificial intelligence in automation. As a result, this incongruity illustrates the important point that the list of quality attributes identified through the literature is suggestive, and not prescriptive. Each chatbot implementation will prioritize different quality attributes at different phases of the system's lifecycle.

**Quality Assessment**

The second question raised in this research is how quality assurance can be performed for chatbots and conversational agents. To do this, we scanned the literature to find six references where other researchers address the question of quality assurance in this domain. Finally, we synthesized the approaches to recommend a composite technique, based on the Analytic Hierarchy Process (AHP) by Saaty (1990).

*Previous Approaches*

The literature review uncovered seven references that focused specifically on quality assurance for chatbots and conversational agents. (Vetter, 2002; Goh et al., 2007; Košir, 2013; Coniam, 2014; Kuligowska, 2015; Meira & Canuto, 2015; Kaleem et al., 2016) Although several other resources were considered, these were the only papers for which quality was the central, unifying theme. Each paper noted the lack of guidance for designing quality into these systems, and evaluating the quality of these systems. Goh called for standard metrics, and Košir bemoaned the lack of them.

Each paper addresses a different aspect of quality assessment, from the effectiveness of responses to individuals questions to the merits of a customized, goal-oriented approach. The main points from each paper are summarized in Table 2.

| Authors | Emphasis | Conclusions & Recommendations |
|---|---|---|
| Vetter (2002) | Constructing test scripts to evaluate utility of | Recommends using PARADISE method (Walker et al., 1997; Sanders & Scholtz, 2002) to determine |

| | conversations | whether conversational constructs meet basic linguistic quality standards. Based on maximizing satisfaction and task success, and minimizing costs. Involves creating a performance function based on confusion matrices (counts of successful and failed communications) for each participant. |
|---|---|---|
| Goh et al. (2007) | Effectiveness of question answering | Precision, Recall, and F1 could be metrics for how well questions are answered, but they fall short. New measures must take into consideration that utility of responses is subjective, different domains have different knowledge repositories, and information is always growing. |
| Košir (2013) | Implementation and assessment of one chatbot (thesis) | To overcome the problems outlined by Goh et al. (2007), use an iterative method and track subjective evaluations from multiple evaluators |
| Coniam (2014) | Examining linguistic quality of chatbot responses | Literature review established that nearly all chatbots met baseline requirements for linguistic accuracy, "grammatical fit", and "meaning fit" suggesting that underlying frameworks and packages are fundamentally sound |
| Kuligowska (2015) | Assessment and comparison of 6 commercial Polish chatbots | By interviews, identified 10 key quality attributes (visual look, speech synthesis, form of interface, basic knowledge, specialized knowledge, conversation abilities, response to unexpected situations, personality traits, personalization options, ability to comment/provide feedback), and evaluated each one on an ordinal scale (1..5) to determine an overall quality assessment (good, very good, excellent) |
| Meira & Canuto (2015) | Determine quality metrics specifically for embodied emotional agents, where affective characteristics dominate | Authors propose a three-level measurement framework (conceptual level goals, operational level goals, and quantitative level goals) that examine quality of architecture and affective quality. Metrics at the quantitative level include cohesion, coupling, size, and communications or services per module (for architecture), and cooperation, likeability, enjoyment, trust, naturalness, reduction of frustration, believability, and interestingness as metrics |
| Kaleem et al. (2016) | Identify and test an assessment approach customized for each conversational agent | The weakness of existing frameworks is that they do not take into account that different conversational systems will have different goals. They adapt the Goal-Question-Metric approach by Fenton & Pfleeger (1998) and suggest pre/post test scores, perception of learning, correct/incorrect responses, and time in system as metrics that could be used for a quality assessment. |

**Table 2.** Quality assessment approaches in previous studies.

The need for a goal-driven approach that incorporates different users' subjective experiences with the chatbots is evident in each of these six results. Furthermore, many useful metrics are suggested, but there is no guidance to determine when each metric should be applied. Although an absolute scale for quality assessment is unlikely given the wide variety of chatbots and conversational agents (and the fact that some are embodied in avatars or humanoid robots), it should be possible to systematically compare the quality of two or more chatbots. If these metrics are linked to individual elements of the comprehensive categorized list of quality attributes presented in Table 1, then Analytic Hierarchy Process (AHP) can be used after a prioritization and down selection process to accomplish this comparison. The following section describes this process.

*Synthesized Approach: Analytic Hierarchy Process (AHP)*

AHP is a structured approach for navigating complex decision-making processes that involve both qualitative and quantitative considerations. First, create a hierarchy of quality attributes and select appropriate metrics to represent each attribute. Second, construct pairwise comparisons between the quality attributes for one or more product options. Next, create comparison matrices and compute the first principal eigenvector of each one to assess relative and global priority. Finally, combine the priorities and compute inconsistency factors to determine which product option best satisfies the hierarchy of quality attributes. (Software is available to make the last two steps easier; code is provided in the Appendix that illustrates how to do this in practice.)

For chatbots and conversational agents, the product options might be 1) two or more versions of the conversational system, or 2) the "as-is" system and one or more "to-be" systems under development. Consider an example where you have made improvements to a chatbot system over the past year, and now you want to see if the quality has improved. First, examine the quality attributes in Table 1 to see which ones are most important, and select metrics (from Table 2, or from your experience). Have multiple users participate in sessions with your chatbots, and record the metrics you selected. An example is shown in Table 3.

| Category | Quality Attribute | Metric | Old | New |
|---|---|---|---|---|
| **Performance** | Robustness to unexpected input | % of successes | 86-92% | 91-93% |
| | Provides appropriate escalation channels | % of successes | 80% | 100% |
| **Humanity** | Transparent to inspection (known chatbot) | % of users who correctly classify | 100% | 100% |
| | Able to maintain themed discussion | 0 (low) .. 100 (high) | 72 (Avg.) 8 (St. Dev.) | 85 (Avg.) 12 (St. Dev.) |

|  | Able to respond to specific questions | % of successes | 68-82% | 80-85% |
|---|---|---|---|---|
| **Affect** | Provides greetings, pleasant personality | 0 (low) .. 100 (high) | 89 (Avg.) 3 (St. Dev.) | 96 (Avg.) 3 (St. Dev.) |
|  | Entertaining, engaging | 0 (low) .. 100 (high) | 50 (Avg.) 21 (St. Dev.) | 66 (Avg.) 4 (St. Dev.) |
| **Accessibility** | Can detect meaning and intent | % of successes | 85-90% | 82-86% |
|  | Responds to social cues appropriately | % of successes | 78% | 77% |

**Table 3.** Example data to collect for an AHP chatbot quality assessment.

AHP uses pairwise assessments between categories, and within categories. The first step is to set up the attribute hierarchy (Figure 3). Notice that the top level displays the goal, the next level shows the quality attribute categories, and the next level from that includes the nine quality attributes that were selected in the prioritization process. At the bottom, the two alternatives (OLD and NEW) are shown. Because the hierarchical structure is central, this approach aligns well with the goal-oriented approach of Kaleem et al. (2016).

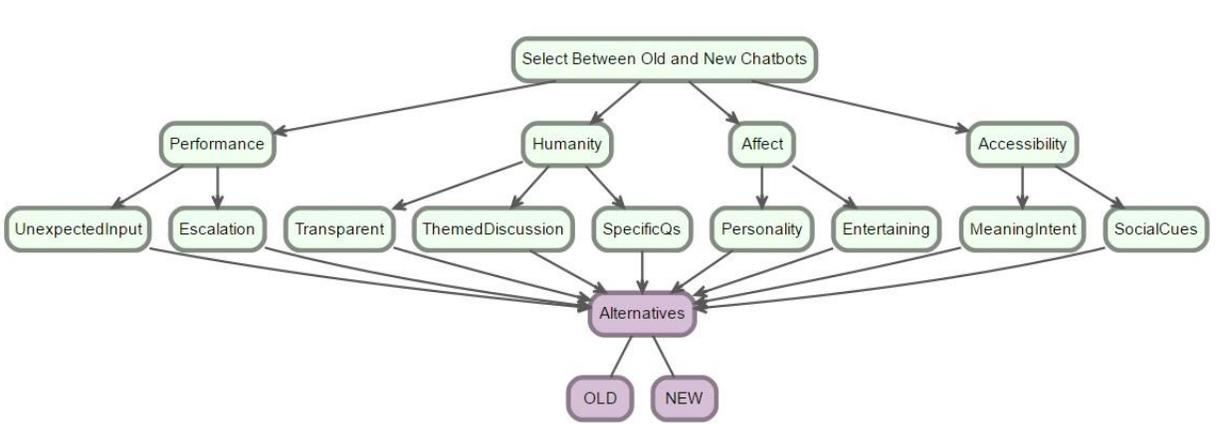

**Figure 3.** Hierarchical model of prioritized quality attributes.

The first step is to make pairwise comparisons between the categories themselves. Typically in AHP, only the numbers 1, 3, 5, 7, and 9 are used. Create a matrix where each cell indicates how much more important the category in the row is as compared to the category in the column. Record the reciprocal of the value if the category in the column is more important than the category in the row. This means that the number 1 will be in the diagonal, since a category cannot be more or less important to itself. For example, in Figure 4, Performance is much more important than Humanity or Affect (7), but is slightly less important than Accessibility (1/3). Values of 9 and 1/9 indicate that there is a substantial difference between the importance of the two categories being assessed.

        Performance   Humanity    Affect    Accessibility

|               |     |     |     |     |
| ------------- | --- | --- | --- | --- |
| Performance   | 1   | 7   | 7   | 1/3 |
| Humanity      | 1/7 | 1   | 1/5 | 1/7 |
| Affect        | 1/7 | 5   | 1   | 1/7 |
| Accessibility | 1/3 | 7   | 7   | 1   |

**Figure 4.** Priority matrix for quality categories.

Within each category, create a priority matrix to express the relative merit of each quality attribute as compared to the others. The "Humanity" attribute was selected for display in Figure 5 because it has the most number of attributes for this example. Notice how each measurement has a complementary cell: for example, Transparent is less important than ThemedDiscussion (1/5), so it makes sense that ThemedDiscussion would be more important than Transparent (5, the reciprocal). There will be four priority matrices total at this level (one for each of the four top-level quality attribute categories).

|                  | Transparent | ThemedDiscussion | SpecificQs |
| ---------------- | ----------- | ---------------- | ---------- |
| Transparent      | 1           | 1/5              | 1/5        |
| ThemedDiscussion | 5           | 1                | 1          |
| SpecificQs       | 5           | 1                | 1          |

**Figure 5.** Priority matrix for selected "Humanity" quality attributes.

Finally, use the measured values (in the "OLD" and "NEW" columns of Table 3) to compare how the different versions of the chatbot perform in terms of each quality attribute. There will be one comparison for each of the nine quality attributes we selected in the original prioritization. Figure 6 shows one of these nine lowest-level priority matrices, showing that the OLD chatbot is much less effective (1/7) in terms of the Escalation attribute - and similarly, the NEW chatbot is much more effective (7).

|     | OLD | NEW |
| --- | --- | --- |
| OLD | 1   | 1/7 |
| NEW | 7   | 1   |

**Figure 6.** Priority matrix for selected "Escalation" quality attribute.

When the first principal eigenvector of each of these matrices are determined and combined, a priority metric for each attribute and category is generated. Although the mechanics of the computations are beyond the scope of this article, instructions are provided in the Appendix for you to use the R Statistical Software (R Core Team, 2016) to perform your own analysis.

When the computations are complete, the results help you choose which alternative best satisfies your quality attributes, given that some attributes are more important, and others are less important. Figure 7 shows the results for this example problem. Unfortunately, our efforts have been in vain: the OLD chatbot is weighted much more heavily (66.2% compared to 33.8%) given the priorities that we set between our categories and attributes.

|  | Weight | OLD | NEW | Consistency |
|---|---|---|---|---|
| Select Between Old and New Chatbots | 100.0% | 66.2% | 33.8% | ⓘ 18.4% |
| Accessibility | 54.5% | 39.1% | 15.3% | 0.0% |
|    MeaningIntent | 47.7% | 35.7% | 11.9% | 0.0% |
|    SocialCues | 6.8% | 3.4% | 3.4% | 0.0% |
| Performance | 32.1% | 24.6% | 7.5% | 0.0% |
|    UnexpectedInput | 28.1% | 21.1% | 7.0% | 0.0% |
|    Escalation | 4.0% | 3.5% | 0.5% | 0.0% |
| Affect | 9.4% | 1.6% | 7.8% | 0.0% |
|    Entertaining | 7.8% | 1.3% | 6.5% | 0.0% |
|    Personality | 1.6% | 0.3% | 1.3% | 0.0% |
| Humanity | 4.1% | 1.0% | 3.1% | 0.0% |
|    SpecificQs | 1.9% | 0.3% | 1.5% | 0.0% |
|    ThemedDiscussion | 1.9% | 0.5% | 1.4% | 0.0% |
|    Transparent | 0.4% | 0.2% | 0.2% | 0.0% |

**Figure 7.** AHP results for the example problem.

The consistency scores should also be reasonable. Ideally these values should be below 10%, but in practice, numbers below 20% are often acceptable. High values in this column indicate some discrepancy in the individual assessments (for example, prioritizing all elements on one level of your hierarchy low with respect to each other). If there is a problem, inspect all of your assessments to make sure that there have been no data input errors.

**Conclusions**

This paper provided a review of the academic literature since 1990, and industry articles since 2015, to 1) gather and articulate quality attributes for chatbots and conversational agents, and 2) discover and synthesize quality assurance approaches to recommend strategies moving forward. There are many ways for practitioners to apply the material in this article:

- The quality attributes (Table 1) can be used as a checklist for a chatbot implementation team to make sure they have addressed key issues

- Two or more conversational systems can be compared by selecting the most significant quality attributes, and/or
- Systems can be compared at two points in time to see if quality has improved, which is particularly useful for adaptive systems that learn as they are exposed to additional participants and topics

The example showed how this goal-oriented approach might be used to evaluate the quality of two different chatbot implementations. Because the method relies on pairwise comparisons, any metric (including those emphasized by the authors in Table 2) can be associated with each quality attribute, and the results will still be valid. Furthermore, this technique can be easily adapted to evaluate different implementations over time, which is essential since most conversational agents learn from experience with users. These factors make the AHP approach particularly robust for assessing the quality of chatbots and conversational agents, resolving the majority of issues identified by previous researchers.

# References

Abran, A., Khelifi, A., Suryn, W., & Seffah, A. (2003, April). Consolidating the ISO usability models. In Proceedings of 11th international software quality management conference (Vol. 2003, pp. 23-25).

Alarifi, A., Alsaleh, M., & Al-Salman, A. (2016). Twitter turing test: Identifying social machines. Information Sciences, 372, 332-346.

Applin, S. A., & Fischer, M. D. (2015, November). New technologies and mixed-use convergence: How humans and algorithms are adapting to each other. In Technology and Society (ISTAS), 2015 IEEE International Symposium on (pp. 1-6). IEEE.

Beilby, L. J., & Zakos, J. (2014). U.S. Patent No. 8,719,200. Washington, DC: U.S. Patent and Trademark Office.

Bhakta, R., Savin-Baden, M., & Tombs, G. (2014, June). Sharing Secrets with Robots?. In EdMedia: World Conference on Educational Media and Technology (Vol. 2014, No. 1, pp. 2295-2301).

Bradeško, L., & Mladenić, D. (2012). A survey of chatbot systems through a loebner prize competition. In Proceedings of Slovenian Language Technologies Society Eighth Conference of Language Technologies (pp. 34-37).

Cohen, D., & Lane, I. (2016, February). An oral exam for measuring a dialog system's capabilities. In Proceedings of the Thirtieth AAAI Conference on Artificial Intelligence (pp. 835-841). AAAI Press.

Coniam, D. (2014). The linguistic accuracy of chatbots: usability from an ESL perspective. Text & Talk, 34(5), 545-567.

Eeuwen, M. (2017). Mobile conversational commerce: messenger chatbots as the next interface between businesses and consumers (Master's thesis, University of Twente). Retrieved on February 1, 2017 from http://essay.utwente.nl/71706/1/van%20Eeuwen_MA_BMS.pdf

Epstein, R. (2007). From Russia, with love. Scientific American Mind, 18(5), 16-17.

Fenton, N. E., & Pfleeger, S. L. (1998). Software Metrics: A Rigorous and Practical Approach: Brooks.

Ferrara, E., Varol, O., Davis, C., Menczer, F., & Flammini, A. (2016). The rise of social bots. Communications of the ACM, 59(7), 96-104.

Ghose, S., & Barua, J. J. (2013, May). Toward the implementation of a topic specific dialogue based natural language chatbot as an undergraduate advisor. In Informatics, Electronics &

Vision (ICIEV), 2013 International Conference on (pp. 1-5). IEEE.

Goh, O. S., Ardil, C., Wong, W., & Fung, C. C. (2007). A black-box approach for response quality evaluation of conversational agent systems. International Journal of Computational Intelligence, 3(3), 195-203.

Hertzum, M., Andersen, H. H., Andersen, V., & Hansen, C. B. (2002). Trust in information sources: seeking information from people, documents, and virtual agents. Interacting with computers, 14(5), 575-599.

Isaac, A. M., & Bridewell, W. (2014). Mindreading deception in dialog. Cognitive Systems Research, 28, 12-19.

Kaleem, M., Alobadi, O., O'Shea, J., & Crockett, K. Framework for the Formulation of Metrics for Conversational Agent Evaluation. (2016). In RE-WOCHAT: Workshop on Collecting and Generating Resources for Chatbots and Conversational Agents-Development and Evaluation Workshop Programme, May 28 (p. 20).

Kerly, A., Hall, P., & Bull, S. (2007). Bringing chatbots into education: Towards natural language negotiation of open learner models. Knowledge-Based Systems, 20(2), 177-185.

Košir, D. (2013). Implementacija in testiranje klepetalnika (Doctoral dissertation, Univerza v Ljubljani).

McElrath, L. (2017, April 2). Watching the hearings, I learned my "Bernie bro" harassers may have been Russian bots. Shareblue. Retrieved on April 2, 2017 from http://shareblue.com/watching-the-hearings-i-learned-my-bernie-bro-harassers-may-have-been-russian-bots/

McTear, M., Callejas, Z., & Griol, D. (2016). Conversational Interfaces: Past and Present. In The Conversational Interface (pp. 51-72). Springer International Publishing.

Meira, M. O., & Canuto, A. M. P. (2015). Evaluation of Emotional Agents' Architectures: an Approach Based on Quality Metrics and the Influence of Emotions on Users. In Proceedings of the World Congress on Engineering (Vol. 1).

Metz, R. (2017, March 23) Three Weeks with a Chatbot and I've Made a New Friend. MIT Technology Review. Retrieved on March 23 from https://www.technologyreview.com/s/603936/three-weeks-with-a-chatbot-and-ive-made-a-new-friend

Miner, A. S., Milstein, A., Schueller, S., Hegde, R., Mangurian, C., & Linos, E. (2016). Smartphone-based conversational agents and responses to questions about mental health, interpersonal violence, and physical health. JAMA internal medicine, 176(5), 619-625.

Morrissey, K., & Kirakowski, J. (2013, July). 'Realness' in Chatbots: Establishing Quantifiable Criteria. In International Conference on Human-Computer Interaction (pp. 87-96). Springer Berlin Heidelberg.

Neff, G., & Nagy, P. (2016). Automation, Algorithms, and Politics| Talking to Bots: Symbiotic Agency and the Case of Tay. International Journal of Communication, 10, 17.

Pauletto, S., Balentine, B., Pidcock, C., Jones, K., Bottaci, L., Aretoulaki, M., ... & Balentine, J. (2013). Exploring expressivity and emotion with artificial voice and speech technologies. Logopedics Phoniatrics Vocology, 38(3), 115-125.

R Core Team (2016). R: A language and environment for statistical computing. R Foundation for Statistical Computing, Vienna, Austria. URL https://www.R-project.org/.

Radziwill, N. M. & Benton, M. C. (2017, March 12). Neurodiversity secrets for innovation and design. SXSW Interactive, Austin TX.

Ramos, R. (2017, February 3) Screw the Turing Test - Chatbots don't need to act human. VentureBeat. Retrieved on March 13, 2017 from https://venturebeat.com/2017/02/03/screw-the-turing-test-chatbots-dont-need-to-act-human/.

Saaty, T. L. (1990). Decision making for leaders: the analytic hierarchy process for decisions in a complex world. RWS publications.

Sanders, G. A., & Scholtz, J. (2001, December). Measurement and evaluation of embodied conversational agents. In Embodied conversational agents (pp. 346-373). MIT Press.

Simonite, T. (2017, March 22). Customer Service Chatbots Are About to Become Frighteningly Realistic. MIT Technology Review, https://www.technologyreview.com/s/603895/customer-service-chatbots-are-about-to-become-frighteningly-realistic/.

Solomon, M. (2017, March 23) If Chatbots Win, Customers Lose, Says Zappos Customer Service Expert. Forbes. Retrieved on March 24, 2017 from https://www.forbes.com/sites/micahsolomon/2017/03/23/customers-lose-if-chatbots-win-says-zappos-customer-service-expert

Staven, T. (2017, March 22). What Makes a Good Bot (or Not)? Unit4 Newsletter. Retrieved on March 23, 2017 from http://www.unit4.com/blog/2017/03/what-makes-a-good-bot-or-not

Storey, M. A., & Zagalsky, A. (2016, November). Disrupting developer productivity one bot at a time. In Proceedings of the 2016 24th ACM SIGSOFT International Symposium on Foundations of Software Engineering (pp. 928-931). ACM.

Taylor, S. J. (2016, Oct 21). Very Human Lessons from Three Brands that Use Chatbots to Talk to Customers. Fast Company. https://www.fastcompany.com/3064845/human-lessons-from-

[brands-using-chatbots](brands-using-chatbots)

Thieltges, A., Schmidt, F., & Hegelich, S. (2016, March). The Devil's Triangle: Ethical Considerations on Developing Bot Detection Methods. In 2016 AAAI Spring Symposium Series.

Thing, V. L., Sloman, M., & Dulay, N. (2007, May). A survey of bots used for distributed denial of service attacks. In IFIP International Information Security Conference (pp. 229-240). Springer US.

Tsvetkova, M., García-Gavilanes, R., Floridi, L., & Yasseri, T. (2016). Even Good Bots Fight. arXiv preprint arXiv:1609.04285.

Turing, A. (1950). Computing machinery and intelligence. Mind, vol. 59, no. 236, pp. 433–460.

Vassos, S., Malliaraki, E., Falco, F. D., Di Maggio, J., Massimetti, M., Nocentini, M. G., & Testa, A. (2016). Art-Bots: Toward Chat-Based Conversational Experiences in Museums. In *Interactive Storytelling: 9th International Conference on Interactive Digital Storytelling, ICIDS 2016, Los Angeles, CA, USA, November 15–18, 2016, Proceedings 9* (pp. 433-437). Springer International Publishing.

Vetter, M. (2002). Quality aspects of bots. In Software quality and software testing in internet times (pp. 165-184). Springer Berlin Heidelberg.

Walker, M. A., Litman, D. J., Kamm, C. A., & Abella, A. (1997, July). PARADISE: A framework for evaluating spoken dialogue agents. In Proceedings of the eighth conference on European chapter of the Association for Computational Linguistics (pp. 271-280). Association for Computational Linguistics.

Wallace R., 2003. The elements of AIML style. ALICE AI Foundation. Retrieved on March 15, 2017 from http://www.alicebot.org/style.pdf

Weiser, M. (1991). The computer for the twenty-first century. *Scientific American*, September, pp. 94–110.

Weizenbaum, J. (1966). ELIZA—a computer program for the study of natural language communication between man and machine. Communications of the ACM, 9(1), 36-45.

Wilson, H. J., Daugherty, P. R., & Morini-Bianzino, N. (2017, March 23) Will AI Create as Many Jobs as it Eliminates? MIT Sloan Management Review. Retrieved on March 24, 2017 from [http://sloanreview.mit.edu/article/will-ai-create-as-many-jobs-as-it-eliminates/](http://sloanreview.mit.edu/article/will-ai-create-as-many-jobs-as-it-eliminates/)

**Appendix: Analytic Hierarchy Process (AHP)**

The `ahp` package in the R Statistical Software (R Core Team, 2016) can be used to process pairwise comparisons data in the manner shown in this paper. If you do not have access to an R installation, you can also do the analysis on the web at https://ipub.com/apps/ahp/ as shown here. There are three steps in the process:

> 1: Prepare a YAML file containing all pairwise comparisons
> 2: Generate and validate the hierarchical model
> 3: Perform the analysis, checking consistency scores and overall weightings of alternatives.

**Step 1: YAML**

YAML (YAML Ain't Markup Language) is a data serialization mechanism that is used in some programming environments. Like FORTRAN, it is sensitive to the column in which each line of text begins. Complete instructions for how to generate a YAML file for AHP can be found at https://cran.r-project.org/web/packages/ahp/vignettes/file-format.html. Here is the YAML file for the example in this article:

```
Version: 2.0

#########################
# Alternatives Section
#
Alternatives: &alternatives
# Here, we list all the alternatives, together with their attributes.
# We can use these attributes later in the file when defining
# preferenceFunctions. The attributes can be quantitative or
# qualitative.
  OLD:
  NEW:
#
# End of Alternatives Section
#####################################

#####################################
# Goal Section
#
Goal:
# The goal spans a tree of criteria and the alternatives
  name: Select Between Old and New Chatbots
  description: >
    Model quality assessment as a decision process.
  author: unknown
  preferences:
    # preferences are typically defined pairwise
    # 1 means: A is equal to B
    # 9 means: A is highly preferable to B
    # 1/9 means: B is highly preferable to A
```

```yaml
    pairwise:
      - [Performance, Humanity, 7]
      - [Performance, Affect, 7]
      - [Performance, Accessibility, 1/3]
      - [Humanity, Affect, 1/5]
      - [Humanity, Accessibility, 1/7]
      - [Affect, Accessibility, 1/7]
  children:
    Performance:
      preferences:
        pairwise:
          - [UnexpectedInput, Escalation, 7]
      children:
        UnexpectedInput:
          preferences:
            pairwise:
              - [OLD, NEW, 3]
          children: *alternatives
        Escalation:
          preferences:
            pairwise:
              - [OLD, NEW, 7]
          children: *alternatives
    Humanity:
      preferences:
        pairwise:
          - [Transparent, ThemedDiscussion, 1/5]
          - [Transparent, SpecificQs, 1/5]
          - [ThemedDiscussion, SpecificQs, 1]
      children:
        Transparent:
          preferences:
            pairwise:
              - [OLD, NEW, 1]
          children: *alternatives
        ThemedDiscussion:
          preferences:
            pairwise:
              - [OLD, NEW, 1/3]
          children: *alternatives
        SpecificQs:
          preferences:
            pairwise:
              - [OLD, NEW, 1/5]
          children: *alternatives
    Affect:
      preferences:
        pairwise:
          - [Personality, Entertaining, 1/5]
      children:
        Personality:
          preferences:
            pairwise:
              - [OLD, NEW, 1/5]
          children: *alternatives
```

```
        Entertaining:
          preferences:
            pairwise:
              - [OLD, NEW, 1/5]
          children: *alternatives
    Accessibility:
      preferences:
        pairwise:
          - [MeaningIntent, SocialCues, 7]
      children:
        MeaningIntent:
          preferences:
            pairwise:
              - [OLD, NEW, 3]
          children: *alternatives
        SocialCues:
          preferences:
            pairwise:
              - [OLD, NEW, 1]
          children: *alternatives
#
# End of Goal Section
######################################
```

**Step 2: Generate and Validate the Hierarchical Model**

Next, go to https://ipub.com/apps/ahp/. Click on the "Model" tab, and copy and paste the entire contents of the YAML file in Step 1 into the text area. *Be sure to start in column 1 of line 1, and be sure not to leave any extra characters or blank lines at the bottom.* To check if the YAML file is valid, click "Visualize". If a picture of your decision hierarchy appears, your YAML file meets the requirements and you may proceed.

**Step 3: Perform the Analysis**

Process your AHP data by clicking "Analyze". Be sure the values in the consistency column are no greater than 20%. The alternative that best meets your goals is the one with the largest weight in the first row.